\documentclass{PoS}

\usepackage{cleveref}
\usepackage{pstricks}
\usepackage{graphicx}
\usepackage{bm}

\newcommand{\beq}{\begin{equation}}
\newcommand{\eeq}{\end{equation}}
\newcommand{\beqs}{\begin{eqnarray}}
\newcommand{\eeqs}{\end{eqnarray}}
\newcommand{\ZvQQ}{\ensuremath{Z_{V^4_{QQ}}}}
\newcommand{\Zvqq}{\ensuremath{Z_{V^4_{qq}}}}

\title{ Leptonic $\bm{B}$- and $\bm{D}$-meson decay constants with 2+1 flavors of asqtad fermions}

\ShortTitle{Leptonic $B$- and $D$-meson decay constants with 2+1 flavors of asqtad fermions}

\author{\speaker{Ethan~T.~Neil} \\
       Department of Physics, University of Colorado, Boulder, CO USA \\
       RIKEN-BNL Research Center, Brookhaven National Laboratory, Upton, NY USA \\
       E-mail: \email{ethan.neil@colorado.edu}}

\author{Andreas~S.~Kronfeld \\
	Fermi National Accelerator Laboratory, Batavia, IL USA \\
	Institute for Advanced Study, Technische Universit\"at M\"unchen, Garching, Germany}

\author{James~N.~Simone \\
	Fermi National Accelerator Laboratory, Batavia, IL USA}

\author{Ruth~S.~Van de Water \\
	Fermi National Accelerator Laboratory, Batavia, IL USA}

\author{(for the Fermilab Lattice and MILC Collaborations)}

\abstract{We present the status of our updated $D$- and $B$-meson decay-constant analysis, based on the MILC $N_f=2+1$ asqtad gauge ensembles. Heavy quarks are incorporated using the Wilson clover action with the Fermilab interpretation. This analysis includes ensembles at five lattice spacings from $a \approx 0.045$ to 0.15 fm, and light sea-quark masses down to 1/20${}^{\rm th}$ of the strange-quark mass.  Projected error budgets for ratios of decay constants, in particular between bottom- and charm-meson decay constants, are presented.}

\FullConference{The 32nd International Symposium on Lattice Field Theory\\
		 23-28 June, 2014\\
		 Columbia University New York, NY}

\begin{document}

\section{Motivation}

The $B$- and $D$-meson leptonic decay constants $f_{B,D}$ are important inputs into the comparison of theory with various experimental results in heavy-quark flavor physics.  The decay rates for fully leptonic decays of $B$ and $D$ mesons are proportional both to the decay constants $f_{B,D}$ and to the appropriate CKM matrix elements; extraction of the latter from experimental results therefore requires a precise determination of the decay constants.

Further, the decay constants are also crucial for the prediction of rates for decay processes that are rare in the standard model, such as $B_s \rightarrow \mu^+ \mu^-$, now being measured to good precision by various LHC experiments \cite{CMS:2014xfa}.  Rare processes such as this can be ideal testing grounds for the standard model, and are potentially quite sensitive to contributions from new physics \cite{Blanke:2014asa}.  Currently, uncertainty in the standard model prediction for this particular process is dominated by theoretical input, including $f_{B_s}$, motivating a precision study of this quantity.

Here we present an update of our ongoing analysis of the MILC asqtad ensembles \cite{Bazavov:2011aa, Neil:2011ku,Bazavov:2013wia} with the goal of extracting the decay constants $f_{B,D}$.  In addition to presenting the current state of the analysis, preliminary error estimates for the ratios $f_B / f_D$ and $f_{B_s} / f_{D_s}$ are presented.  In combination with other precise numerical results for the $D$-meson decay constants (such as with HISQ charm quarks \cite{Bazavov:2014wgs}), these may be used for improved determinations of the $B$-meson decay constants.

\section{Simulation details}

We employ the MILC gauge-field ensembles with asqtad staggered fermions. The gauge ensembles are generated with $2+1$ flavors of dynamical sea quarks, and a tadpole-improved gauge action.  Heavy quarks are included using the clover fermion action with the Fermilab heavy-quark interpretation.  For further details on the lattice action and simulation parameters, see \cite{Bailey:2014tva}.

\section{Analysis of correlation functions}

We compute heavy-light two-point correlation functions with both smeared and local sources, and both pseudoscalar and axial-vector interpolating operators.  For a correlation function with source type $i$ and sink type $j$, the expected functional form factorizes as
\begin{equation}
C_{ij}(t) = \sum_{n=0}^{N_X} \left[ A_{i,n} A_{j,n} \left( e^{-E_n t} + e^{-E_n (N_t - t)} \right) - (-1)^t A_{i,n}' A_{j,n}' \left( e^{-E_n' t} +e^{-E_n' (N_t - t)} \right) \right]
\end{equation}
where $N_t$ is the temporal extent of the lattice and $N_X$ denotes the number of excited states included.  Making use of time-reversal symmetry, all correlators are considered only in the range $0 \leq t \leq N_t/2$, with the results for $t > N_t/2$ averaged in before the analysis is done.  The decay constant is determined from the ground-state amplitude of the pseudoscalar-axial vector correlation function, obtained from a joint fit to four correlators: pseudoscalar-pseudoscalar with both sources local or smeared, and pseudoscalar-axial vector with both local and smeared pseudoscalar sources.

For each two-point function, a corresponding fit range $[t_{\rm min}, t_{\rm max}]$ is chosen.  We use $t_{\rm min}/a = 4$ for all fits, and increase $N_X$ until an acceptable correlated fit is obtained; this requires $N_X = 3$ on most ensembles, with $N_X = 4$ used for the ensembles with the largest $N_t$.  $t_{\rm max}$ is held constant in physical units, approximately $1.5$ fm for fits to bottom-quark correlators, and $2.0$ fm for charm.  If this constraint gives $t_{\rm max}/a > 30$, we set $t_{\rm max}/a = 30$ instead to avoid fit covariance matrices which are poorly determined by the available statistics; because the signal-to-noise ratio for the correlators drops quickly at large $t$, this is not expected to significantly affect the final fit results.

With a large number of excited states included, several techniques are used to ensure the numerical stability of the fits.  For all manifestly positive fit parameters, the logarithm of the parameter is used in the minimization in order to impose the positivity constraint on the fitter.  In any multi-exponential fit, a potential ambiguity in the ordering of the energy states $E_0, E_1, E_2, ...$ results in the existence of multiple equivalent minima in the space of all fit parameters.  Rather than using the energies directly as fit parameters, we fit to the set
\beq
E_0,\ \log(E_1 - E_0),\ \log(E_2 - E_1), ...
\eeq
which imposes the explicit ordering $E_0 < E_1 < E_2 < ...$ on the fit function.

Finally, we use empirical Bayesian constraints \cite{Lepage:2001ym} for all correlator fits.  Priors imposed on the excited states are relatively loose, and are used only for numerical stability.  For the ground-state parameters,  we employ a ``two-stage" fitting procedure.  In the first stage, a single-exponential fit is carried out restricted to data in a region corresponding roughly to an observed ``plateau" in the effective mass $m_{\rm eff}(t)$; the plateau is taken from $t_{\rm max}$ as given above, up to approximately 2.2 fm for $B$-meson fits, and 3.0 fm for $D$-meson fits.

\begin{figure}
\centering
\includegraphics[width=0.8\textwidth]{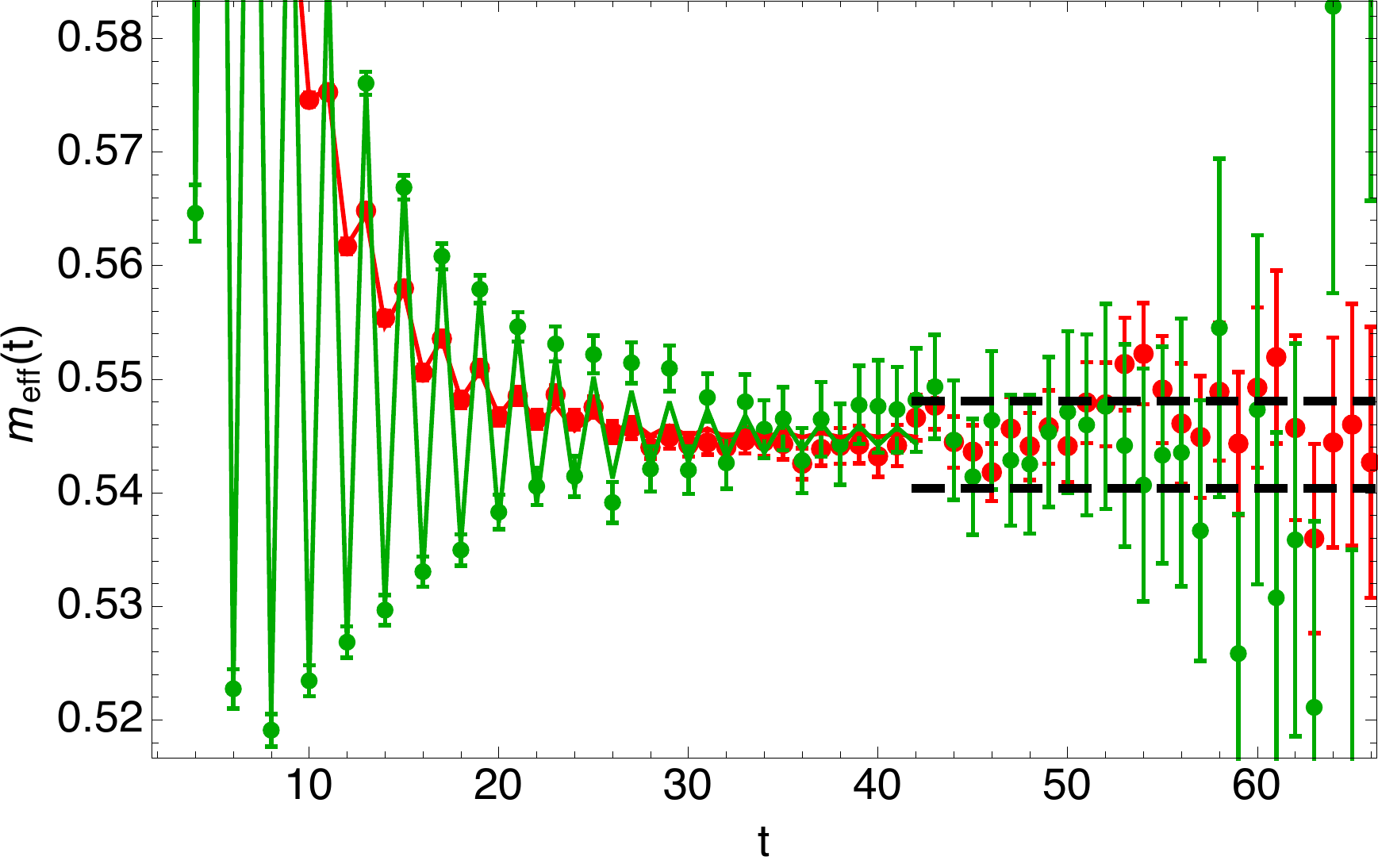}
\caption{Results of the ``two-stage" fit procedure described in the text, for correlators with a local pseudoscalar source and either local pseudoscalar sink (red) or local axial-vector sink (green), shown as an effective mass.  The band (black, dashed lines) shows the uncertainty used in the empirical Bayesian constraint for the full fit, obtained from the ``plateau" fit result for $E_0$ as described in the text.  The solid lines joining the effective mass points are reconstructed from the full fit, and show good agreement to low $t$.  \label{fig:corr}}
\end{figure}

This first-stage fit is then used to set Bayesian prior constraints for the second-stage fit, which are carried out to the range $[t_{\rm min}, t_{\rm max}]$ as described above.  The prior mean values for ground-state parameters are set to the first-stage best-fit values.  Prior widths are set to three times the one-sigma classical errors on the parameters from the first stage.  The results of this procedure are shown for a representative correlator are shown in \Cref{fig:corr}.

\begin{figure}
\centering
\includegraphics[width=0.99\textwidth]{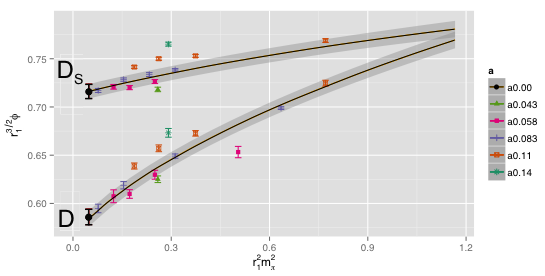}
\includegraphics[width=0.95\textwidth]{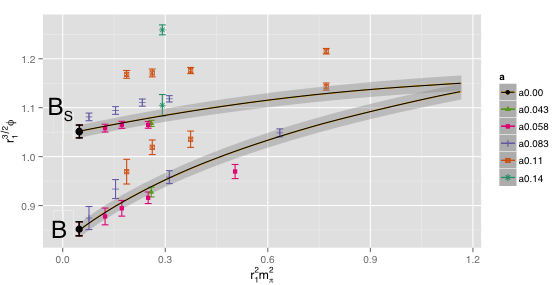}
\caption{Preliminary, blinded results for the chiral-continuum extrapolation of the decay ampitude $r_1^{3/2} \phi_{H_q}$ for the $D$ (upper panel) and $B$ (lower panel) heavy-light mesons. Only full-QCD points (valence and sea light-quark masses equal) are shown in the plot, but the fit includes all available partially-quenched data.  \label{fig:chiral}}
\end{figure}

\section{Chiral-continuum extrapolation}
Using the results of the two-point fits and separate calculations of renormalization constants, we construct the renormalized decay amplitude $\phi_{H_q}$ from the ground-state amplitude associated with the axial-vector sink:
\beq
r_1^{3/2} \phi_{H_q} = (r_1/a)^{3/2} \sqrt{2 Z_{V^4_{qq}} Z_{V^4_{QQ}} } \rho_{A^4_{Qq}} A_{A^4_{Qq},0},
\eeq
where $q$ and $Q$ are the light and heavy quark making up heavy-light meson $H_q$, respectively.  This quantity is proportional to the decay constant times the square root of the meson mass, $\phi_{H_q} = f_{H_q} \sqrt{M_{H_q}}$.

The chiral-continuum extrapolation for $\phi_{H_q}$ is carried out separately for $B$ and $D$ mesons using the framework of partially quenched heavy-light meson staggered chiral perturbation theory (PQHMs$\chi$PT) \cite{Bazavov:2011aa}.  With the increased precision of this updated analysis, we find that to obtain acceptable chiral fits, we must include next-to-next-to-next-to-leading order (N${}^3$LO) analytic terms with cubic dependence on the light-quark masses.

Our preliminary chiral-continuum fits are shown in \Cref{fig:chiral}.  We emphasize that these are preliminary, blinded results, and include an unknown multiplicative factor.  The overall goodness of fit is equal to $\chi^2 / \textrm{dof} = 99.4/85$ ($p$-value: 0.14) for the $D$-meson system, and $\chi^2 / \textrm{dof} = 78.1/85$ ($p$-value: 0.69) for the $B$-meson system.  

\section{Preliminary error budget for decay-constant ratios}

\begin{table}
\begin{center}
	\caption{Projected error budgets for the heavy-flavor decay-constant ratios.  All errors are quoted as percentages.  Error projections are preliminary, and may change when estimated from the full chiral fit in the final analysis.  Entries marked with a dash show quantities that do not contribute due to cancellation in the given ratios.  \label{tab:err}}
	\begin{tabular}{lllll}
	\hline
	\hline
	Source & $f_{D_s}/f_{D}$ & $f_{B_s} / f_B$ & $f_{B} / f_D$ & $f_{B_s} / f_{D_s}$  \\
        \hline
        Statistics & 0.2\%  & 0.4\% & 0.6\% & 0.5\% \\
        Heavy-quark discretization & 0.7\% & 0.6\% & 0.8\% & 0.8\% \\
        Light-quark discretization & 0.2\% & 0.1\% & 0.3\% & 0.3\% \\
        Chiral extrapolation & 0.6\% & 0.6\% & 1.0\% & 0.8\% \\
        Heavy-quark tuning & 0.1\% & 0.1\% & 1.7\% & 1.8\% \\
        $\Zvqq$ & 0.0\% & 0.0\% & --- & --- \\
        $\ZvQQ$ & ---  &  --- & 0.2\% & 0.2\% \\
        Finite volume & 0.2\% & 0.2\% & 0.3\% & 0.3\% \\
        Higher-order $\rho_{A^4_{Qq}}$ & 0.1\% & 0.1\% & 4.1\% & 4.1\% \\	
        \hline
        Total projected error & 1.1\% & 0.9\% & 4.7\% & 4.7\%\\
        \hline
        \hline
        \end{tabular}
\end{center}
\end{table}

Of particular interest in this calculation are various ratios of heavy-light decay constants, since these quantities can be calculated with better precision than the individual decay constants due to the cancellation of various systematic effects.  In our previous work \cite{Bazavov:2011aa}, the ratios $f_{D_s} / f_D$ and $f_{B_s} / f_B$ were reported.  In this updated analysis, we plan to also include results for the bottom-to-charm ratios $f_{B_s} / f_{D_s}$ and $f_B / f_D$.  The combination of these ratios with more precise results for charm decay constants \cite{Bazavov:2014wgs} will enable improved determinations of the bottom decay constants.  An estimated preliminary error budget for each ratio is shown in \Cref{tab:err}; the procedure for estimating each uncertainty contribution is detailed below.

We first consider systematic errors due to the renormalization.  The lattice axial-vector current is renormalized with the combined factor
\begin{equation}
Z_{A^4_{Qq}} = \rho_{A^4_{Qq}} \sqrt{Z_{V^4_{QQ}} Z_{V^4_{qq}} }.
\end{equation}
In the heavy-flavor ratios $f_{B_s} / f_{D_s}$ and $f_B / f_D$, the dependence on $Z_{V^4_{qq}}$ cancels, so only the ratios $\sqrt{Z_{V^4_{bb}}/Z_{V^4_{cc}}}$ and $\rho_{A^4_{bq}}/\rho_{A^4_{cq}}$ contribute to the error.  On the other hand, for the ratios $f_{D_s} / f_D$ and $f_{B_s} / f_B$ the factor $Z_{V^4_{QQ}}$ cancels; dependence on the ratio $Z_{V^4_{ss}} / Z_{V^4_{qq}}$ remains, but because this quantity -- which formally differs from 1 at order $\alpha_s m_s^2 a^2$ --  is found to have vary slowly with respect to the valence quark mass, so the overall contribution to the systematic error is expected to be negligible.

For the ratios $f_{B_s} / f_{B}$ and $f_{D_s} / f_D$, effects of higher-order terms in the perturbative factor $\rho_{A^4_{Qq}}$ enter only through the small variation of the perturbative renormalization factors with the valence light-quark mass; we retain the estimated error of $0.1\%$ in the ratios as derived in \cite{Bazavov:2011aa}.  In the ratios $f_{B} / f_D$ and $f_{B_s} / f_{D_s}$, the light-quark valence mass dependence cancels, but dependence on the heavy-quark mass, which is expected to be significant, remains.  We conservatively ignore correlations between the two $\rho$-factors appearing in these ratios, and estimate the projected systematic error by varying numerator and denominator by the value $\rho_{\rm max}^{[1]} \alpha_s$, which $\rho_{\rm max}^{[1]} = 0.1$ and $\alpha_s = \alpha_V(2/a)$ evaluated with $a \approx 0.045$ fm (see \cite{Bailey:2014tva} for notation.)

Heavy-quark discretization errors are estimated via power-counting using the explicit formulas in Appendix A of \cite{Bazavov:2011aa}.  Six such terms contribute at $\mathcal{O}(a^2)$ and $\mathcal{O}(\alpha_s a)$; each term is multiplied by an unknown fit parameter, for which nominal values are taken from the preliminary joint chiral-continuum extrapolation shown above.

Another source of uncertainty in the bottom-to-charm ratios is from the (mis)tuning of the heavy-quark masses; this error cancels almost entirely in the ratios of decay constants with the same heavy quark.  To estimate the heavy-quark tuning error, we model the dependence of $\phi_{H_q}$ on the heavy-quark mass $M_Q$ by including the leading $1/M_Q$ dependence expected from heavy-quark effective theory,
\beq
\phi_{H_q}(M_Q) = \phi_\infty - \frac{\phi'_\infty}{M_Q}
\eeq
The parameters $\phi_\infty$ and $\phi'_\infty$ are estimated by comparing nominal values of $\phi_{H_q}$ for bottom and charm from our previous complete analysis \cite{Bazavov:2011aa}.  Fractional uncertainties in the heavy-quark masses are estimated to be $1.8\%$ for charm and $2.7\%$ for bottom based on uncertainties in $\kappa$ reported in \cite{Bailey:2014tva}, and are then used with classical error-propagation formulas to estimate the uncertainty in the decay constants, assuming a correlation coefficient of $1$ for the heavy-quark mass appearing in $M_{B}$ and $M_{B_s}$, and similarly for the charm mesons.

Light-quark discretization errors are estimated to scale as $\mathcal{O}(\alpha_s a^2)$, with improvement based on the reduction of the minimum available lattice spacing from $a \sim 0.09$ fm to $a \sim 0.045$ fm.  Chiral extrapolation errors are estimated by taking the estimated errors from \cite{Bazavov:2011aa} and projecting them to reduce according to the lightest available light-quark mass in $r_1$ units, $r_1 m_q$.   Statistical errors are projected to improve as $\sqrt{N_{\rm cfg}}$ (the number of sources used on all configurations except the coarsest is held constant at 8.)  In terms of $m_\pi L$, the largest volumes available are essentially the same as in the previous work \cite{Bazavov:2011aa}, so the finite-volume error is estimated to be unchanged. 

Adding the individual errors in quadrature, this leads to projected total errors on the percent level for the ratios $f_{D_s} / f_D$ and $f_{B_s} / f_B$, and total error of $4.7\%$ for the bottom-to-charm ratios $f_{B} / f_D$ and $f_{B_s} / f_{D_s}$.  The latter ratios, in combination with more precise determinations of the charm decay constants \cite{Bazavov:2014wgs}, could lead to values for $f_B$ and $f_{B_s}$ with precision comparable to other state-of-the-art calculations \cite{Christ:2014uea,Bernardoni:2014fva,Bussone:2014bya,Aoki:2014nga} if the uncertainty due to higher-order perturbative renormalization can be reduced.

\paragraph{Acknowledgments}
This work was supported by the U.S. Department of Energy, the National Science Foundation, and the URA Visiting Scholars' Program, and by the German Excellence Initiative and the European Union Seventh Framework Programme under grant agreement No.~291763 as well as the European Union's Marie Curie COFUND program (A.S.K.). Fermilab and BNL are operated under contracts DE-AC02-07CH11359 and DE-AC02-98CH10886 respectively, with the DOE. Computations were carried out at the Argonne Leadership Computing Facility, the National Center for Atmospheric Research, the National Center for Supercomputing Resources, the National Energy Resources Supercomputing Center, the National Institute for Computational Sciences, the Texas Advanced Computing Center, and the USQCD facilities at Fermilab, under grants from the NSF and DOE.

\end{document}